\documentstyle[12pt]{article}
\begin{document}
\noindent
\begin{center}
{\LARGE {\bf Mechanism for a\\ Decaying Cosmological Constant
\\}}
\vspace{2cm}
${\bf Y.~ Bisabr}$ \footnote{e-mail:~y-bisabr@cc.sbu.ac.ir.}~,~
${\bf H.~ Salehi}$ \footnote{e-mail:~h-salehi@cc.sbu.ac.ir.}\\
\vspace{0.5cm}
{\small {Department of Physics, Shahid Beheshti University, Evin,
Tehran 19839,  Iran.}}\\
\end{center}
\vspace{1cm}
\begin{abstract}
A mechanism is introduced to reduce a large cosmological constant to a 
sufficiently small value consistent with the observational upper
limit.  The basic ingradient in this mechanism is a distinction which has
been made between the two unit systems used on cosmology and particle
physics.  We have used a conformal invariant gravitational model to define a
particular conformal frame in terms of the
large scale properties of the universe.  It
is then argued that the contributions of mass scales in particle physics
to the vacuum energy density should be considered in a different conformal
frame.  In this manner a cancellation mechanism is presented in which
the conformal factor plays a key role to relax the large effective
cosmological constant.
\end{abstract}
\vspace{3cm}
Cosmological observations imply that there exists an extremely small upper
limit on the vacuum energy density in the present state of our
universe.  This stands in sharp contradiction with theoretical
predictions \cite{1}.  In fact, any mass scale 
in particle physics contributes to the
vacuum energy density much larger than this upper bound \cite{2}.  It 
therefore
seems that a, yet unknown, cancellation mechanism must take place to
reduce these huge contributions to a sufficiently small value.  One may
expect that such a mechanism 
has two characterictics.  Firstly, it should
not be sensitive to a particular type of contribution and should work equally 
well for every mass scale introduced by particle physics.  Secondly, it
should work whenever these contributions are considered at cosmological
level since the discrepancy manifests when one compares them with 
relevant cosmological observations.  This latter strongly suggests
that construction of a mechanism for relaxing these contributions 
should somehow take into account
the distinction between the two unit systems defined by cosmology
and particle physics.  One should note that the small 
upper limit is obtained in a unit system which is defined in terms
of large scale cosmological parameters.  On the other hand, the 
theoretical predictions
are based on a natural unit system which is suggested
by quantum physics.  With this fact in mind, the point we wish
to make here is that the 
large descrepancy, which is usually known as the cosmological constant
problem, arises when one prejudices that the two unit systems should be
indistinguishable up to a constant conversion factor in all spacetime
points.  It means that they should transform to each other by a global
unit transformation.  Such a global transformation clearly carries no 
dynamical implications and the use of a particular unit system is actually
a matter of convenience.  It means that 
one may arbitrarily use the unit system suggested 
by quantum physics to describe the
evolution of the universe or the cosmological unit system to describe
dynamical properties of an elementary particle.\\
In this Letter we introduce a theoretical scheme in which an explicit
recognition is given to the distinguished characteristics of these
two unit systems.  In such a theoretical shceme one should no
longer accept the triviality one usually asigns to a unit
tranformation.  In this respect, we shall consider
local unit transformations, conformal
transformations \cite{3}, which relates different standard of 
units, conformal frames, via general spacetime dependent 
conversion (conformal) factors.  In this language 
the observational and the theoretical predictions on the
vacuum energy density are actually carried out in two different 
conformal frames.  We emphasize that local unit 
transformations give a dynamical
meaning to changes of unit systems and can 
be consequently taken as a basis for constructing a dynamical
mechanism which works due to cosmic expansion.  Along this 
line of investigation, we shall show that the conformal factor
which relates the two unit systems plays the role of a dynamical field
which can eventually reduce the effective cosmological constant to a
small value consistent with observations.\\
To begin with, let us cosider a gravitational model described
by the action\footnote{Throughout this paper we work in units
in which $\hbar=c=1$ and the sign conventions are 
those of MTW \cite{5}.} \cite{4}
\begin{equation}
S=\frac{1}{2} \int d^{4}x \sqrt{-g}~ (g^{\alpha\beta} \nabla_{\alpha}\phi
\nabla_{\beta}\phi+\frac{1}{6} R \phi^{2})+S_{m}(g_{\alpha\beta})
\label{1}\end{equation}
in which $\nabla_{\alpha}$ denotes a covariant 
differentiation, $\phi$ is a scalar
field, $R$ is the Ricci scalar and $S_{m}(g_{\alpha\beta})$ is the action of
some matter fields.  The gravitational part of this action remains
invariant under conformal transformations
\begin{equation}
\bar{g}_{\alpha\beta} =e^{2\sigma}  g_{\alpha\beta}
\label{2}\end{equation}
\begin{equation}
\bar{\phi} = e^{-\sigma}  \phi
\label{3}\end{equation}
where $\sigma$ is a smooth dimensionless spacetime function.  Variation 
with
respect to $g^{\alpha \beta}$ and $\phi$ yields, respectively,
\begin{equation}
G_{\alpha\beta}=6\phi^{-2}(T_{\alpha\beta}+\tau_{\alpha\beta})
\label{5}\end{equation}
\begin{equation}
\Box \phi-\frac{1}{6}R \phi=0
\label{6}\end{equation}
where
\begin{equation}
\tau_{\alpha\beta}=-(\nabla_{\alpha }
\phi \nabla_{\beta}\phi-\frac{1}{2}g_{\alpha \beta} \nabla_{\gamma  }\phi
\nabla^{\gamma  }\phi)+\frac{1}{6}(\nabla_{\alpha }\nabla_{\beta}
-g_{\alpha\beta}\Box
)\phi^2
\end{equation}
\begin{equation}
T_{\alpha \beta}=\frac{-2}{\sqrt{-g}}
\frac{\delta  }
{\delta g^{\alpha\beta} } S_{m}({g}_{\alpha \beta})
\label{8}\end{equation}
Here $\Box \equiv g^{\alpha \beta} \nabla_{\alpha }
\nabla_{\beta}$ and $G_{\alpha \beta}$ is the Einstein tensor.  If one substitutes
the equation (\ref{6}) into the trace of (\ref{5}) one
obtains $T^{\gamma}_{\gamma}=0$ which means that the 
stress tensor of matter must be traceless.  This is a consequence
of the absence of a dimensional parameter in the gravitational
part of the action (\ref{1}).  One may consider matter systems
with nonvanishing trace by introducing a term such as
\begin{equation}
-\frac{1}{6} \int d^{4} x \sqrt{-g} \mu^2 \phi^2
\label{x8}\end{equation}
in which $\mu$ is a 
parameter with dimension of mass.  In this case, one
obtains 
\begin{equation}
G_{\alpha\beta}+\mu^2 g_{\alpha\beta}=6\phi^{-2}(T_{\alpha\beta}
+\tau_{\alpha\beta})
\label{5'}\end{equation}
\begin{equation}
\Box \phi-\frac{1}{6}R \phi+\frac{1}{3} \mu^{2} \phi=0
\label{6'}\end{equation}
and
\begin{equation}
T^{\gamma}_{\gamma}=\frac{1}{3} \mu^2 \phi^2  
\label{x3}\end{equation}
The parameter $\mu$ also allows us to study breakdown of the
conformal invariance in this model.  In general, in a conformal transformation
all dimensional parameters
are required to be transformed according to their dimensions so
that $\mu$ should obey the
transformation rule $\mu \rightarrow e^{-\sigma} \mu$.  The 
conformal invariance can, however, be broken when a particular
conformal frame is chosen in which the dimensional parameter 
takes on a constant 
configuration.  The choice of such a specific 
conformal frame, or equivalently
the constant value attributed to $\mu$, is merely
suggested by the physical conditions one wishes to investigate
in the problem at hand.  In the conformal frame
which is introduced in this way the 
gravitational coupling is then characterized
by the preferred configuration of $\phi^{-2}$. \\ 
Here we first intend to determine a conformal frame with use of the
large scale properties of the observed universe.  In this case, the
most suggestive choice for $\mu$
is $\mu \sim H_{0}$ with $H_{0}^{-1}$ being the present 
Hubble radius of the universe.  In 
the corresponding conformal
frame, which is referred from now on as the cosmological frame, 
we try then to find the preferred constant configuration 
of $\phi^{-2}$ in terms of large scale distribution of matter
contained in the universe.  To do this, we take $T_{\alpha\beta}$ to be 
a pressureless perfect fluid with energy density $\rho$ and
write the gravitational equations (\ref{5'}) for a spatially flat
Friedman-Robertson-Walker metric
\begin{equation}
ds^2=-dt^2+a(t)(dx^2+dy^2+dz^2)
\label{frw}\end{equation}
where $a(t)$ is the scale factor.  In the present epoch of
evolution of the universe this leads to
\begin{equation}
H_{0}^{2} \sim \rho_{0}\phi^{-2}
\label{qq}\end{equation}
in which $\rho_{0}$ is the present energy density of 
cosmic matter.  We also
use the observational fact that $\rho_{0} \sim \rho_{c}$ which
$\rho_{c} =\frac{3H_{0}^{2}}{8\pi G}$ is 
the critical density.  The relation (\ref{qq}) then 
gives 
\begin{equation}
\phi^{-2} \sim G
\end{equation}
Thus in the
cosmological frame $\phi^{-2}$ takes a constant configuration which is given
by the gravitational constant.  By
implication the action (\ref{1}), together with (\ref{x8}), reduces to 
\begin{equation}
S=\frac{1}{16\pi G} \int d^{4}x \sqrt{-g} (R-2\mu^2)+  
S_{m}(g_{\alpha\beta})
\label{x4}\end{equation}
This
corresponds to the usual Einstein-Hilbert action with a cosmological
constant induced due to finite size of the 
universe.\\
The cosmological constant, however, receives strong contributions
from various
mass scales introduced by particle physics. To
clarify how these contributions
enter our gravitational model in the cosmological frame, it should 
be remarked that the  
conformal invariance breaking might be considered as a result of
emergence of rest masses of elemetary particles in appropriately small
energy scales.  One could therefore define an alternative conformal frame
in which these masses take some constant values.  Such a conformal frame would
evidently have properties entirely different 
from those that used to define the cosmological
frame.  In particular, it should be defined in terms 
of local characteristics
of a typical elementary particle and should neglect the large scale properties
of the universe.  It follows that when 
one compares contributions of the rest
masses to vacuum energy density with relevant cosmological
observations 
one should note the fact that they belong to two 
different conformal frames.  This means that 
the rest masses should have a variable contribution to 
vacuum energy density in the cosmological frame.  We 
would like to analze this assertion
by writing the action (\ref{x4}) in the form
\begin{equation}
S=\frac{1}{16\pi G} \int d^{4}x \sqrt{-g} \{R -2(\mu^2
+\bar{\Lambda}) \}+S_{m}(g_{\alpha\beta}) 
\end{equation}
where $\bar{\Lambda}=e^{-2\sigma} \Lambda$ and $\Lambda$ being
a typical mass scale in particle physics. Since the conformal factor 
appears here as a new dynamical
degree of freedom we let the above action involve a kinetic
term for $\sigma$ so that
\begin{equation}
S=\frac{1}{16\pi G} \int d^{4}x \sqrt{-g} 
\{R-2(\mu^2+\Lambda e^{-2\sigma})
-\epsilon g^{\alpha\beta}\nabla_{\alpha}
\sigma \nabla_{\beta}\sigma\}+S_{m}(g_{\alpha \beta})
\label{4}\end{equation}
where $\epsilon$ is a dimensionless constant parameter of
order of unity.  In this
way for $\epsilon >0$, $\sigma$ acts as a normal dynamical field 
with positive energy density which allows us to
investigate the evolution of $\bar{\Lambda}$ as the universe evolves.  We now 
follow the consequences of the action (\ref{4}) by
writing the field equations 
\begin{equation}
G_{\alpha\beta}+(\mu^2 +\Lambda e^{-2\sigma})g_{\alpha\beta}
-\epsilon (\nabla_{\alpha}\sigma \nabla_{\beta}\sigma-\frac{1}{2}g_{\alpha\beta} 
\nabla^{\gamma}\sigma \nabla_{\gamma}\sigma)
=8\pi G T_{\alpha\beta}
\label{f}\end{equation}
\begin{equation}
\Box \sigma+\frac{2}{\epsilon} \Lambda e^{-2\sigma}=0
\label{ff}\end{equation}
In these equations the exponential coefficient for $\Lambda$ emphasizes
the distinction between the two unit systems mentioned above.  In an
expanding universe this distinction is expected to
increase since cosmological scales enlarge as the universe expands and
the conformal factor $e^{2\sigma}$ must grow with time.  This authomatically
provides us with a dynamical reduction of $\bar{\Lambda}$ in
the cosmological frame.  That this intuitive picture is actually
consistent with the field equations is illustrated in
the following:\\
Applying (\ref{f}) and (\ref{ff}) to the metric (\ref{frw}), yields
\begin{equation}
3 H^2-\Lambda_{eff}=8\pi G \rho
\label{13}\end{equation}
\begin{equation}
\dot{\rho}+3H \rho=0
\label{r}\end{equation}
\begin{equation}
\ddot{\sigma}+3H \dot{\sigma}-\frac{2}{\epsilon} \Lambda e^{-2\sigma}=0
\label{15}\end{equation}
with
\begin{equation}
\Lambda_{eff}=\mu^{2}+\Lambda e^{-2\sigma}+\frac{1}{2} \epsilon \dot{\sigma}^2
\label{16}\end{equation}
where $H=\frac{\dot{a}}{a}$ and the overdot indicates 
differentiation with respect to $t$.  Due to 
homogeneity and isotropy, the field $\sigma$ is
taken to be only a fuction of time.\\
The equation (\ref{r}) immediately gives
\begin{equation}
\rho a^3 =const.
\label{rh}\end{equation}
Assuming that the universe follows a power law expansion, namely
that $H \sim t^{-1}$, we use the ansatz
\begin{equation}
e^{\sigma}=\sigma_{0} t
\end{equation}
which satisfies the equation (\ref{15}) with
$\sigma_{0} \sim \sqrt{\Lambda}$.  From (\ref{16}) one then
obtains $\Lambda_{eff} \sim t^{-2}$  which gives
the observational upper limit.  With this result, the 
equation (\ref{13}) suggests that
$\rho \sim t^{-2}$ which is consistent with  
evolution of matter
density in the standard cosmological model.\\
In summary, it 
is argued that different contributions to the energy
content of vacuum may come from different conformal frames.  In 
particular, we have shown that the rest masses
of elementary particles should be regarded as contributions coming from
a conformal frame with properties different from those that suggested
by cosmological observations.  This line of reasoning then leads us to
introduce a mechanism in which the conformal factor appears
as a dynamical field and makes a large cosmological constant damp during
evolution of the universe.  It should be remarked that the nontrivial
behaviour of the conformal factor in such a mechanism may have
notable outcomes at early universe.  For instance, it is worthwhile
to study the effect of such a dynamical field on the inflationary
scenarios or on the possible avoidance of the cosmological
singularity.  We intend to investigate these issues elsewhere.

\end{document}